# Advancing Cloud Computing Capabilities on gem5 by Implementing the RISC-V Hypervisor Extension

George-Marios Fragkoulis Nikos Karystinos George Papadimitriou Dimitris Gizopoulos
University of Athens, Greece
{gm.fragkoulis, n.karystinos, georgepap, dgizop}@di.uoa.gr

## Abstract

This paper presents the implementation and evaluation of the H (hypervisor) extension for the RISC-V instruction set architecture (ISA) on top of the gem5 microarchitectural simulator. The RISC-V ISA, known for its simplicity and modularity, has seen widespread adoption in various computing domains. The H extension aims to enhance RISC-V's capabilities for cloud computing and virtualization. In this paper, we present the architectural integration of the H extension into gem5, an open-source, modular platform for computer system architecture research. We detail the modifications required in gem5's CPU models and virtualization support to accommodate the H extension. We also present evaluation results regarding the performance impact and functional correctness of the extension's implementation on gem5. This study not only provides a pathway for further research and development of RISC-V extensions but also contributes valuable insights into the optimization of the gem5 simulator for advanced architectural features.

## 1 Introduction

The open standard, simplicity, and modular design of the RISC-V instruction set architecture (ISA) have led to widespread attention in the computing industry [19]. RISC-V is a foundation that can be used for many applications, such as embedded systems, high-performance computing (HPC), and cloud computing [29]. One of its most significant features is its extensibility, which allows the integration of custom extensions to meet specific computational needs [6, 7, 16]. A notable example of this is the H (Hypervisor) extension, recently introduced and ratified to enhance RISC-V's capabilities in virtualization and cloud computing. The demand for these paradigms is driving the need for more effective and adaptable microprocessor designs. Multiple virtual machines (VMs) can be operated on a single physical machine to optimize resource usage and provide flexibility in managing workloads. To achieve effective virtualization, hardware assistance is required to minimize overhead and enhance performance [17]. These demands are met by the H extension in the RISC-V ISA [6], which provides hardware support for hypervisor-level operations. Several factors motivated the implementation and open-sourcing of the H extension within the gem5 simulation framework, as enumerated below:

1. Researchers can explore the potential of RISC-V for virtualization and high-performance computing by incorporating the H extension into gem5. This implementation provides the foundation for further innovations and improvements within the RISC-V ecosystem.
2. The simulation of the H extension in gem5 facilitates comprehensive performance evaluation and benchmarking. This helps researchers understand how the extension affects system performance, especially in virtualized environments.
3. The educational potential of gem5 is widely utilized for teaching computer architecture. The inclusion of the H extension in gem5 provides students with a valuable educational tool for analyzing advanced architectural features and their ramifications.
4. As an open-source project, gem5 gets contributions from a global community of developers and researchers. The implementation of the H extension promotes collaboration and knowledge sharing, thereby fostering innovation in the field of computer architecture.

The main purpose of this paper is to examine how the H extension is implemented in the gem5 simulator [1, 14, 22], a widely used tool in computer architecture research. The primary contributions of this paper are as follows:

(1) We implement and integrate the H extension into the latest version of the gem5 simulator (v24.0), detailing the modifications required in CPU models and virtualization support.
(2) We assess the performance and functional correctness of the H extension implementation in gem5 with and without running VMs.
(3) We provide insights into the impact of the H extension on RISC-V's performance in virtualized environments to guide further research and development.

# 2 Background

## 2.1 The gem5 simulator & the H extension

gem5 [1, 14, 22] is a state-of-the-art, open-source, cycle-level, full-system simulator that can provide accurate performance evaluation results and supports many commercial CPUs ISAs. gem5 flexibly provides support for configuring a rich set of microarchitectural parameters (number, size, organization of cores, pipelines, caches, buffers, queues, speculation structures, etc.). The gem5 simulator can operate with the functional (atomic) and the cycle-level (detailed) microarchitecture detail. Moreover, it can work by performing syscall emulation (SE) or full system (FS) simulation. In this paper, we are based on the FS mode and the atomic CPU of gem5 for the porting and evaluation of the H extension on top of RISC-V CPUs. Also, we present how to extend the gem5 simulator to provide a complete open-source detailed system simulation framework that incorporates that hypervisor extension for accelerated evaluation of RISC-V software deployments.

The H extension is a RISC-V ISA extension that can enhance virtual memory support in virtualized environments. This extension can improve performance and security by using hardware to perform virtual address translation and manage page tables, reducing overhead in virtualization. This is especially beneficial in cloud computing and data centers, where virtualization is extensively used to enhance resource utilization and decrease expenses. Moreover, the H extension can enforce memory protection and isolation in virtualized environments, preventing unauthorized memory access and improving security. Additionally, it can enhance the efficiency of virtual memory management, improving performance and reducing power consumption. The H extension is widely supported and utilized in many industries, including cloud computing, data centers, and embedded systems [25].

Figure 1 shows the required features for the development of the RISC-V Hypervisor extension. It provides a standardized interface for hypervisors (i.e., Xvisor [23], KVM [24]) to access and control the VMs running on the system. Specifically, the H-extension requires the following key features:

(1) It should define new instructions and registers for managing virtual machine state and virtual interrupts.
(2) It should provide support for memory virtualization and device virtualization to enable efficient sharing of hardware resources among multiple virtual machines.
(3) It should provide support for multiple privilege levels, allowing hypervisors to have higher privileges than the guest operating systems they manage. The base ISA includes three privilege levels: M (Machine), which has the highest privileges; S (Supervisor), used for system-level operations; and U (User), utilized for user applications. With the H extension enabled, a new privilege level called HS (Hypervisor-extended Supervisor) is introduced. This level operates similarly to the S level but includes additional registers. In a guest context, there are two additional modes: VS (guest's OS) and VU (guest's applications). Hence, having the H extension enabled, the privilege levels in decreasing order of accessibility are M (Machine), HS, VS, and VU.
(4) It should ensure isolation between VMs by providing mechanisms for enforcing access control policies and preventing unauthorized access to VM resources.

## 2.2 Xvisor bare-metal hypervisor

Xvisor [2] is an open-source type-1 (bare-metal) hypervisor designed to offer a lightweight, portable, and flexible virtualization solution. It delivers high performance and a low memory footprint for various CPU architectures, including RISC-V. The Xvisor code is highly portable, making it adaptable to most general-purpose architectures. It primarily supports full virtualization, accommodating a broad range of unmodified guest operating systems, including Linux [10]. It includes many features typical of modern hypervisors, such as device tree-based configuration, a threading framework, runtime loadable modules, dynamic guest creation/destruction, network virtualization, input device virtualization, etc.

# 3 Implementation Methodology

Implementing the H-extension in the gem5 simulator to support virtualization in the RISC-V ISA requires extensive changes in core parts of the simulator. The main parts of this extension can be described in three sections: *1. Registers*, *2. Exceptions & Interrupts Handling* and *3. Two-Stage*

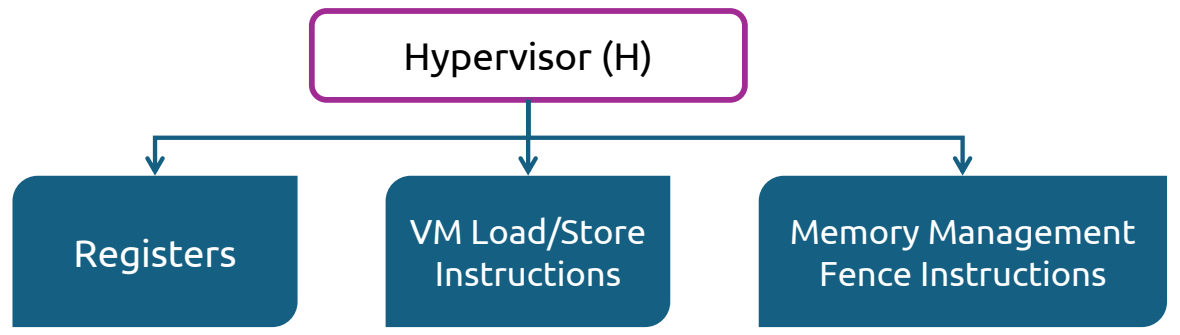

**Figure 1: Illustration of the Hypervisor (H) extension.**

*Translation*. Each section encompasses new definitions described in the RISC-V ISA specification [7], which are added to specific header files in gem5, as well as modifications to the core logic, such as handling faults or page table walking. After implementing the aforementioned parts of the H-extension in gem5, we present the way we test the implementation with various scenarios, including comparing the architectural state under complex conditions, such as handling exceptions based on the values of new hypervisor registers. Further testing involved booting a Type-1 hypervisor, named Xvisor [2, 3] in gem5, and evaluating its behavior by running benchmarks in both host and guest OS.

## 3.1 Registers

Control and Status Registers (CSRs) are responsible for encoding permissions related to memory access and instruction execution at a certain privilege level. The hypervisor extension adds more CSRs to manage two-stage address translation and control the behavior of a VS-mode guest as shown in Table 1. Additionally, manipulating CSRs must respect the read-only permissions of certain bit fields and maintain privilege protection among the registers, meaning some registers cannot be accessed in lower privilege modes.

Moreover, CSRs have bit-fields that serve as aliases for other CSRs. This means that reading or modifying one CSR also involves accessing parts of another CSR. For example, reading the *HVIP* (Hypervisor Virtual Interrupt Pending) CSR includes reading the *MIP* (Machine Interrupt Pending) CSR because the *VSSIP* (Virtual Supervisor Software Interrupt Pending) bit of *HVIP* is an alias of the *VSSIP* bit in *MIP*. Further, accessing supervisor CSRs in VS mode is modified so that access is redirected to the virtual supervisor registers instead (see Table 1). gem5 defines the CSRs in the file *arch/riscv/misc.hh*. Some CSRs are mapped to the same hardware register, such as *SSTATUS* and *MSTATUS*. To prevent access to certain bit fields from lower privilege levels, the simulator utilizes READ REGISTERS MASKS. We extend this approach by adding WRITE REGISTERS MASKS to ensure that read-only bits remain unchanged. Also, register swapping in VS-mode and the new write masks added in *arch/riscv/standard.hh::CSRExecute()* which implements the main functionality of the CSR-managing instructions.

**Table 1: Overview of the implemented registers.**

| | |
|---|---|
| mstatus | *mpv* and *gva* fields added. *mpv* stores the previous virtualization when a trap is taken to M mode and *gva* is written when a trap happens to a guest virtual address and taken to M mode |
| hstatus | H-extension CSR that manages the exception handling behavior of a VS mode guest |
| mideleg | New read-only 1-bit fields for VS and guest external interrupts have been introduced, meaning these interrupts are now handled by HS mode |
| hideleg<br>hedeleg | H-extension CSR that handles the delegation of VS interrupts and traps to VS mode |
| mip, mie | New bit fields for hypervisor interrupts |
| hvip | H-extension CSR that allows a hypervisor to signal virtual interrupts intended for VS mode |
| hip<br>hie | H-extension CSRs for VS-level and hypervisor-specific interrupts |
| hgeip, hgeie | H-extension CSRs for guest external interrupts |
| hcounteren | H-extension CSR for accessing the HPM the virtual machine |
| htval | H-extension CSR for storing the guest physical address that faulted, shifted right by 2 bits when the fault is handled by HS mode |
| mtval2 | H-extension CSR for storing the guest physical address that faulted, shifted right by 2 bits when the fault is handled by M mode |
| hgatp | H-extension CSR that is responsible for the second stage of a guest virtual address translation hgatp holds the PPN of the guest physical root table |
| vsstatus<br>vsip, vsie, vstvec<br>vsscratch, vsepc<br>vscause<br>vstval, vstap | H-extension CSRs that are used in place of the supervisor CSRs when virtualization mode is enabled |

## 3.2 Exceptions & Interrupts Handling

The H extension defines new interrupts and exceptions handled differently based on the current privilege level and the values of the delegation registers. gem5 utilizes the function *RiscvFault::invoke()* for handling both exceptions and interrupts. This is responsible for modifying the status and cause registers. It also calculates the new program counter value and the privilege level. New cases have been added for delegating faults at VS level, including handling faults such as Virtual Instruction Fault. Interrupt detection has been modified to account for the new register for interrupt pending and the new bit fields in the status register.

Figure 2 presents an example of an interrupt handling in atomic (functional) CPU type of the gem5. Assume that the interrupt should be delegated to HS level. In every tick (the gem5-specific unit of simulation time), the CPU calls *Check-Interrupts()*, which reads the *interrupt pending* and *enable* registers, as well as the *delegation* registers based on the current privilege level (*mideleg* is read if the current privilege is lower than M, and *hideleg* is read if the current privilege is lower than HS). If an interrupt is detected, a fault is created and handled by a specific interrupt handler according to the values of the aforementioned CSRs. After the interrupt handler completes, execution may resume.

## 3.3 Two-Stage Address Translation

The translation process of a virtual address is described in the ISA specification [7]. The Sv39 scheme, which utilizes 39

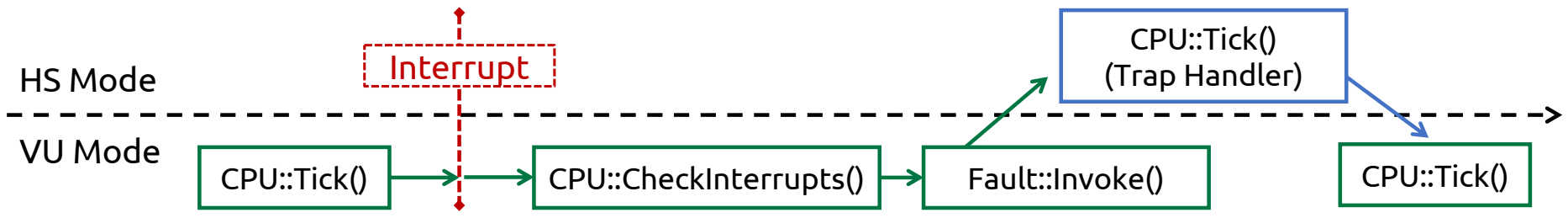

**Figure 2: The gem5's interrupt handling behavior.**

bits of a 64-bit virtual address to provide a three-level page table hierarchy, is shown in Figure 3. In this scheme, the virtual address is divided into several fields: a 25-bit unused field, three 9-bit Virtual Page Number (VPN) fields ($VPN_2$, $VPN_1$, and $VPN_0$), and a 12-bit page offset (assuming a 4KB page size in this example). Each $VPN_i$ serves as an index into the corresponding level of the page table. The base address of the page table is a physical address, and the sum of this base address and the $VPN_i$ offset yields the address of the next-level page table or the final physical page. Guest address translation consists of two stages, with each stage involving a full-page table walk, as shown in Figure 3. However, in virtualized environments, the base register (satp) is split into two different (hypervisor) registers, the *vsatp* and the *hgatp*, supporting address translation in both guest and host layers. The first stage, known as the *VS-Stage*, is controlled by the *vsatp* register and is responsible for translating a guest virtual address to a guest physical address, which is also a virtual address. Subsequently, this guest physical address must be translated to a host physical address. This process, called *G-Stage* translation, utilizes the *hgatp* register. Both the *vsatp* and *hgatp* registers hold the base address of the appropriate page table and the current translation mode, such as Sv39 and Sv39x4 respectively (the guest physical address is widened by 2 bits). Thus, every page table address is virtual and must be translated to a physical address by the G-stage.

The gem5 walking process starts by initializing basic features like the translation mode (gem5 supports only Sv39) and storing some notable fields of status registers, which are used to identify invalid accesses that caused page faults. Subsequently, *arch/riscv/pagetablewalker.hh::walk()* performs the walk by calling *arch/riscv/pagetablewalker.hh::stepWalk()* for intermediate page table accesses. To support the two-stage address translation, the *walk()* procedure is redesigned. This procedure calculates the intermediate addresses for the guest page table and calls *walkGStage()* for G-Stage translation.

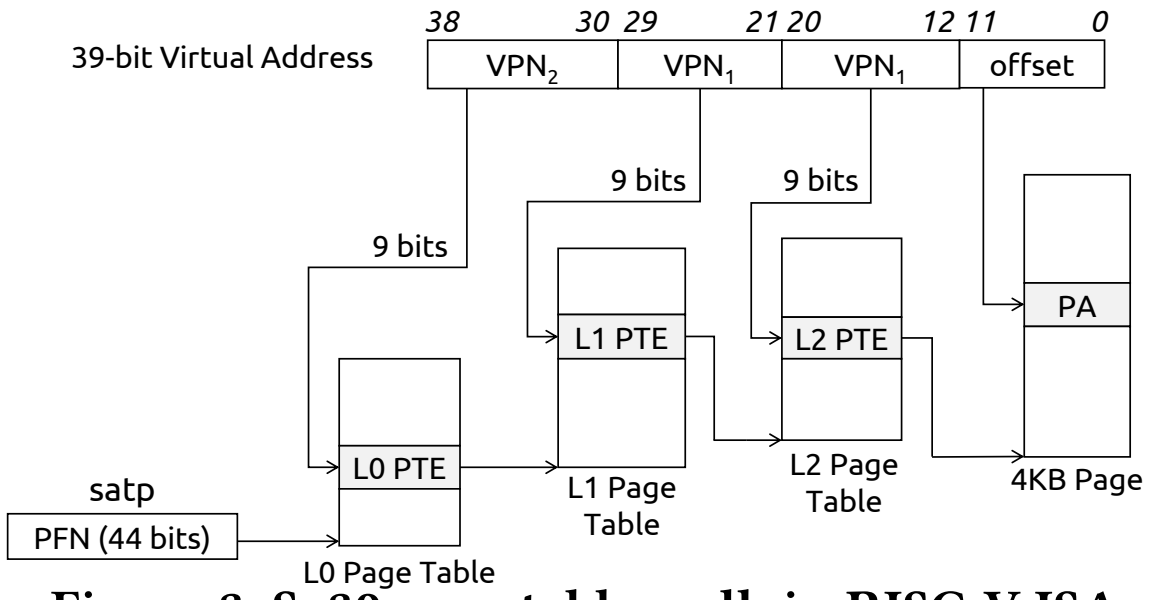

**Figure 3: Sv39 page table walk in RISC-V ISA.**

Subsequently, if no fault is raised, the *stepWalk()* is called to access the guest page table. This process is repeated until a valid page is found, or a page fault occurs due to missing permissions. New page fault conditions, such as *Load Guest Page Fault*, have been added to *arch/riscv/tlb.hh::checkPermissions()* incorporating notable attributes of this access, such as the current privilege mode. In addition, we have implemented new memory instructions that access memory as if virtualization mode is on, utilizing the newly defined *ArchFlagsType* for RISC-V in *arc/riscv/memflags.hh::XlateFlags*. These flags contain information about the instruction, such as forced virtualization, the *HLVX* option (a hypervisor load requiring execute permission), and the *LR* option (load reserved), which has been added to the *arc/riscv/isa/decoder.isa* file. Furthermore, templates describing the behavior of hypervisor loads and stores have been added to *arc/riscv/isa/formats/mem.isa*.

## 3.4 Validation

We thoroughly validate our implementation using specialized tests [4]. They compare the system's behavior to the expected outcomes in various scenarios. Specifically:

- *tinst_tests*: These tests check the *tinst* value written after a page fault. This value, defined by the specification, can be either zero, an instruction trapped to accelerate hardware by skipping the loading of the trapping instruction from memory, or a specific pseudoinstruction encoding to provide additional information about the guest page fault.
- *wfi_exception_tests*: These tests validate the conditions under which faults occur from the *wfi* instruction or verify the successful execution of the instruction.
- *hfence_tests*: Execute *hfence* instructions affecting only the guest TLB (Translation Lookaside Buffer) entries.
- *virtual_instruction*: These tests perform various instructions that, under certain conditions, result in a virtual instruction fault.
- *interrupt_tests*: These tests write to interrupt pending and enable registers and check the cause affected by the interrupt priority [5] and the privilege level that handled the interrupt.
- *check_xip_regs*: These tests validate the aliasing of the interrupt pending registers and the encryption of some of their bit fields because, at some privilege levels, there is no access to information from higher privilege levels.
- *m_and_hs_using_vs_access*: These tests execute hypervisor load and store instructions, validating the data

from memory or the page fault caused by page permissions or the *mstatus* value.

- *second_stage_only_translation*: Performs *G-Stage* translations only by setting the *vsatp*'s mode to zero (BARE).
- *two_stage_translation*: Validates the complete two-stage translation by checking the final translation or throwing a page fault with the correct information (code, privilege mode handled, *gva*, and *tval2* values).

### 3.5 Challenges

Initially, we utilized a Linux kernel packaged with the Berkeley Bootloader (bbl). Unfortunately, bbl requires the *MIDELEG* register, which handles interrupt delegation, to be set to 0x222 (indicating that supervisor software, time, and external interrupts should be delegated). The issue arises because the H extension mandates that virtual interrupts (software, time, and external) must also be delegated from Machine Mode, resulting in an incorrect condition that prevents Linux from booting. Consequently, we opted to use the latest version of gem5 and the SBI bootloader [8]. After implementing the critical parts of H-extension, we boot Xvisor. During this attempt, we encountered several challenges, including: (1) Default creation of the device tree by gem5 caused I/O issues. We modified the device tree based on Spike's device tree [9], (2) The H extension affects many common instructions, including floating-point operations. Access to the Floating-Point Unit (FPU) is controlled by the float status bits in status CSRs. Consequently, when virtualization mode is enabled, the *vsstatus* should also be checked, and (3) The necessity of TLB modification arises from its ability to bypass the page table walking procedure by storing the translation of a virtual page number (VPN) to a physical frame number (PFN), along with some permission bits. Due to the two-stage translation, it is crucial to store both the guest PFN and supervisor PFN to effectively support megapage or gigapage translation. Additionally, it is necessary to store the permission bits of the guest page table entry in gem5's TLB because, in virtualization mode, the guest assumes that the physical address is derived from the guest PFN, which may have different permissions than the supervisor PFN.

## 4 Experimental Results

We present results from nine workloads of the MiBench Suite [20], executing with and without the hypervisor extension enabled (i.e., with and without a VM). All executions are performed on an AMD EPYC 7402 CPU.

### 4.1 Simulation Time Overhead

We first present the time overhead induced by the guest OS when executing a workload. Note that the Linux boot time is 10 times longer when running in a VM compared to native execution in gem5. Therefore, every benchmark simulation we present in this section utilizes gem5's checkpoint functionality to ensure that only the current benchmark is being studied. Additionally, some benchmarks exhibit significantly lower simulation times than others, making it difficult to see their bars in the presented graphs. Figure 4 shows the simulation times in seconds for benchmarks running natively in a full system setup of the gem5 (green bars) versus running in a guest VM in the same environment (red bars). The blue horizontal line indicates the slowdown of each benchmark between the guest OS and the host OS. With an average slowdown of 50%, these benchmarks show an increase in execution time ranging from approximately 30% to 100%.

### 4.2 Executed Instructions Overhead

The hypervisor creates virtual representations of physical hardware resources for VMs. This process, which includes managing virtual CPUs and virtual network interfaces, introduces some computational overhead and results in additional executed instructions, as shown in Figure 5. In this figure, a notable difference in the number of executed instructions is presented. Xvisor operates directly on the hardware and oversees the guest operating system (Linux in our case). Consequently, several factors cause the execution of the same program on a guest OS to involve slightly more instructions than running it directly on the hardware [28]. One major reason is the additional executed hypervisor instructions. The hypervisor executes additional instructions for managing the VM, such as scheduling, and resource allocation. Moreover, certain instructions or operations performed by the guest OS or applications within the VM might need to be trapped by the hypervisor and emulated. For example, privileged instructions that cannot be executed directly by the guest OS must be handled by the hypervisor, adding extra instructions to the execution path. Another reason is the memory management. As we described earlier, virtualized environments use a layer of memory management (i.e., two-stage address translation [11, 13]). Managing these additional layers of memory translation requires extra instructions compared to the single-layer memory management in a native run. Finally, additional instructions need operations intercepted by the hypervisor that require elevated privileges, such as setting hardware registers, configuring interrupts, or accessing certain control registers.

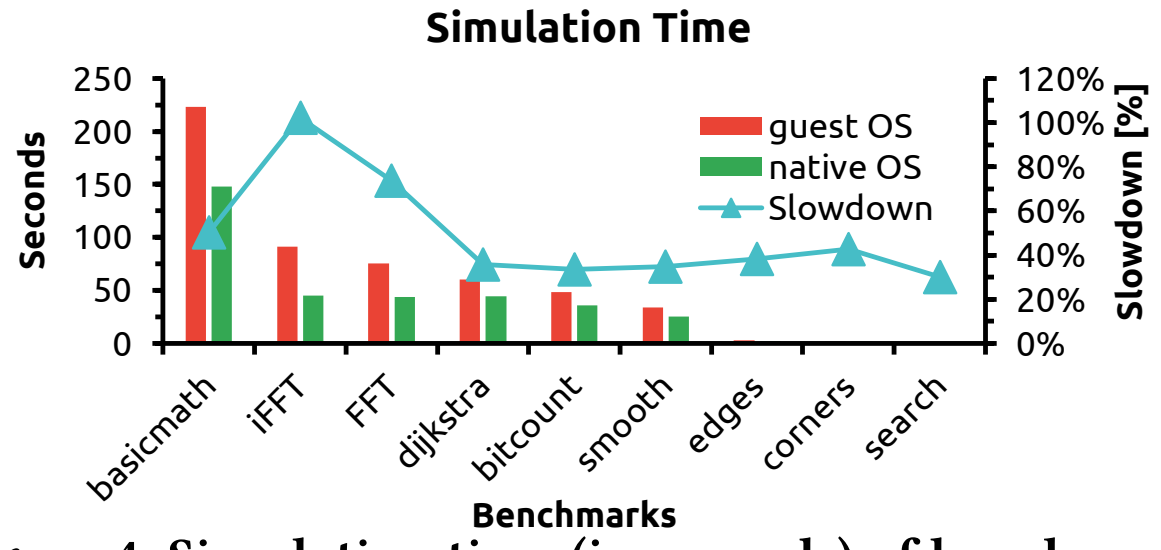

**Figure 4: Simulation time (in seconds) of benchmarks between native and guest OS executions.**

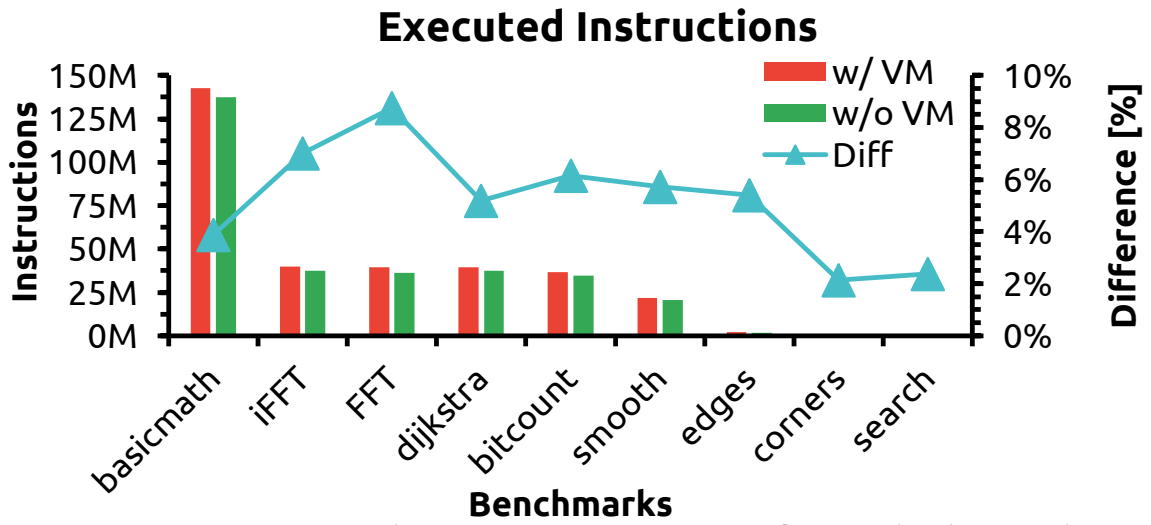

**Figure 5: Executed instructions of each benchmark running with (w/) or without (w/o) VM (i.e., a guest OS).**

## 4.3 Exceptions handled overhead

Figures 6, 7 show the exception handling in the native OS and the guest OS, respectively. Each benchmark includes colored bars representing the number of exceptions handled at specific privilege levels. In the native OS, exceptions can be delegated to two possible levels (M and S), whereas in the guest OS, there are three possible levels (M, HS, and VS) due to the hypervisor extension. When an exception occurs, the delegation registers are read to determine the appropriate privilege level for handling. The two-stage translation process in the guest OS involves more accesses compared to the base translation, leading to a higher frequency of page faults. This trend is evident in the results presented in Figures 6, 7. Additionally, it is noteworthy that the number of exceptions delegated to the S level in the native OS and the VS level in the guest OS are nearly equal, highlighting a similar distribution of exception handling across these levels.

# 5 Related Work

The RISC-V H extension is a relatively new addition to the privileged architecture of the RISC-V ISA, ratified in December '21 [7]. This extension opens up numerous opportunities for both academia and industry to explore advanced virtualization capabilities within the RISC-V ecosystem. Many projects have been undertaken to integrate the hypervisor extension into various models. From a hardware perspective, the Rocket core [26], the CVA6 core [27], and the Legarto core [18] fully support the H-extension. In addition to hardware, software tools, and simulators have also adopted the hypervisor extension. For instance, QEMU [12], a widely used open-source emulator, and the official RISC-V ISA simulator, Spike [9], have both integrated support for the H extension. Furthermore, gem5 [22] already supports full system (FS) simulation [21] enabling a Linux kernel boot for RISC-V ISA. Also, many microarchitectural components complement RISC-V CPUs, such as accelerators [15].

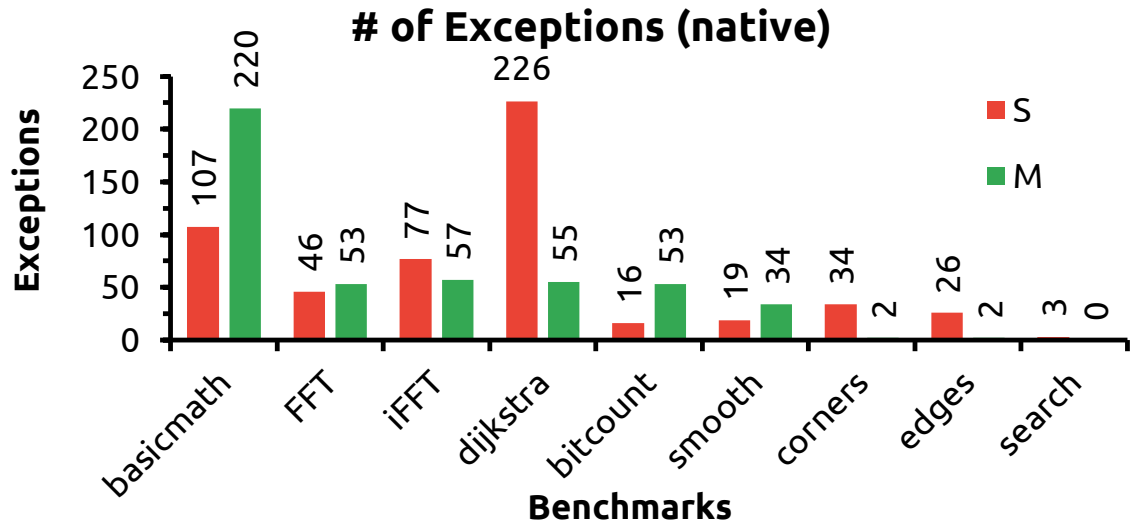

**Figure 6: Number of exceptions for native execution and the privilege levels at which they are delegated.**

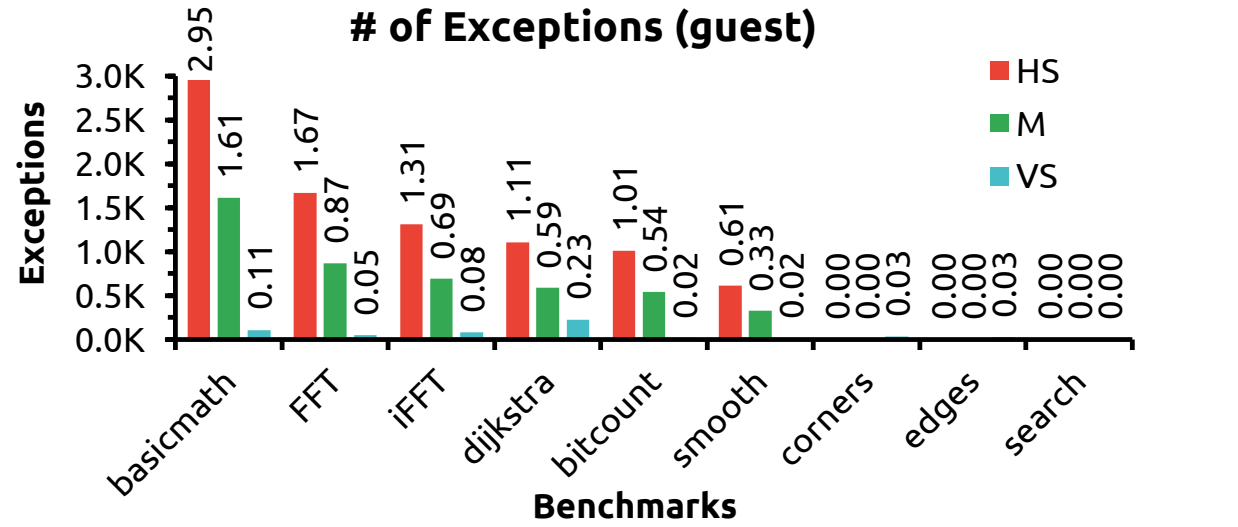

**Figure 7: Number of exceptions handled by the guest OS and the privilege levels at which they are delegated.**

# 6 Conclusion & Future Work

We explored the H (Hypervisor) extension for the RISC-V ISA and implemented it in the gem5 simulator. This integration demonstrates the cloud computing capabilities it brings to the cycle-level simulator. We discussed our modifications to gem5's simulator and its virtualization support, with benchmarking results confirming the functional correctness of the H extension. This work aims to encourage further innovation in computer system architecture using RISC-V and gem5. Future work includes extending support to all ISA-compliant virtual address sizes, all CPU types in gem5, and KVM (a Type-2 hypervisor). We do not expect major disruptions, as gem5 already supports full-system simulation and the hypervisor extension. Furthermore, we plan to use our open-source implementation to enable comprehensive microarchitectural design space exploration for cloud deployments.

# Acknowledgments

This work was supported by the EU's Horizon Europe research and innovation programme under grant agreements No 101093062 (Vitamin-V), No 101097224 (REBECCA), and No 101070238 (NEUROPULS). Views and opinions expressed are however, those of the authors only and do not necessarily reflect those of the EU. Neither the European Union nor the granting authority can be held responsible for them. This project is also carried out within the framework of the National Recovery and Resilience Plan Greece 2.0, funded by the European Union– NextGenerationEU (Implementation body: HFRI) with the title "Reliable Highly Parallel Systems by Design (REDESIGN) and Project Number 16973.